\documentclass[aps,superscriptaddress]{revtex4-2}
\usepackage{amsmath,amssymb}
\usepackage{graphics,graphicx}
\usepackage{subfigure}
\usepackage{dcolumn,bm}
\usepackage{psfrag}
\usepackage{color}
\usepackage{adjustbox}
\usepackage{url}
\usepackage{eqnarray,amsmath}
\usepackage{braket}
\newcommand{\cu}
{\affiliation{Vidyasagar College, 39 Sankar Ghosh lane, Kolkata 700006, India.}}
\newcommand{\be}
{\begin{equation}}
\newcommand{\ee}
{\end{equation}}

\begin{document}
\title{QUANTUM WALKER IN PRESENCE OF A MOVING DETECTOR}
\author{Md Aquib Molla}
\cu
\author{Sanchari Goswami}
\cu
\date{\today}

\begin{abstract}
    In this work, we study the effect of a moving detector on a discrete time one dimensional Quantum Random Walk where the movement is realized in the form of hopping/shifts. The occupation probability $f(x,t;n,s)$ is estimated as the number of detection $n$ and amount of shift $s$ vary. It is seen that the occupation probability at the initial position $x_D$ of the detector is enhanced when $n$ is small which is a quantum mechanical effect but decreases when $n$ is large. The ratio of occupation probabilities of our walk to that of an Infinite walk shows a scaling behavior of $\frac{x_D^2}{n^2}$. It shows a definite scaling behavior with amount of shifts $s$ also.  The limiting behaviors of the walk are observed when $x_D$ is large, $n$ is large and $s$ is large and the walker for these cases approach the Infinite Walk, The Semi Infinite Walk and the Quenched Quantum Walk respectively. \\ \\

\end{abstract}

\maketitle

\section{INTRODUCTION}
Discrete time \textbf{Quantum Random Walk} (QRW) on a line caught the attention of scientists at the end of twentieth century. 
The term quantum walk, the analogue of classical walk was first coined by Aharonov et. al. in 1993 \cite{Aharonov}. Several studies have been made in this regard to find out the differences between quantum walk and the classical walk \cite{Nayak, Kempe, Ambainis}. The quantum interference in QRW results in $\langle x^2 \rangle \propto t^2$ where $x$ is position of the walker and $t$ is the time. Therefore the QRW is quadratically faster compared to the Classical Random Walk (CRW). The situation becomes interesting when quenching phenomena is studied for 
QRW. Slow quenching 
are usually studied in the aspect of  
spin glass system where there are many minima in the energy landscape, separated 
by barriers which may be overcome by quantum tunnelling \cite{Das1, Das2}. Fast quenching is applied for ultracold atoms in an optical lattice
by shifting the position of the trap potential and studying its response \cite{combined2}. In certain systems 
having a quantum critical point, fast or slow quenching of a few variables are used to study nonequilibrium dynamics \cite{Combined1}.
 Quenching in QRW 
is studied earlier in \cite{goswami} where a detector put initially in the path of the QRW was withdrawn after a certain time. 
While experimental study of a QRW is becoming more and more important in recent years, the role of a detector in its path 
is to be studied minutely. The present study is entirely focused to the QRW with a detector in its path.\\

In this work, we studied a QRW with a detector which is not fixed in its position for 
long but can move. Here the movement is taken in form of hopping over the sites. 
The situation can also be understood in some other way from the point of view of an experimental study. The earliest experimental studies can be found in \cite{Schreiber1, Schreiber2}. In \cite{Schreiber1}, it was concluded that the detector must have some role in the discrepancies between the experiment and theoretical model, as there is dead time for the detector. In \cite{photon} also, the dead time of detectors have been addressed and to upgrade the results, use of shorter dead time detectors has been recommended. This may be related to the efficiency of the detector, which may be assumed to be decreasing with time. If such things happen, we may replace the detector 
by another one.  In our case the new detector is placed at some other position with a certain systematic. The same behavior would be observed if a detector stays at a particular position and then hops over to another site.\\

In section II, we present the exact scheme to study QRW in presence of a moving detector. In the same section, we define a few related notations we will use throughout the next sections of this work. Section III consists of the main results of our study. In section IV, a brief summary as obtained from the results is presented.
\section{SCHEME TO STUDY QRW IN PRESENCE OF A MOVING DETECTOR}
The QRW is drastically different from a Classical Random Walk (CRW). For QRW, there is an 
additional degree of freedom called ``Chirality". The chirality here
can take two states ``left" $\ket{L}$ or ``right" $\ket{R}$ and is coupled to the position. The state of the walker is therefore expressed in $\ket{x}\bigotimes\ket{d}$ basis, 
where $\ket{x}$, $\ket{d}$ are the position and chirality eigenstates respectively. 
The wavefunction at position $x$ for time $t$ is,\\
\begin{equation}\label{eq1}
\Psi(x,t) = \begin{pmatrix}
\psi_L(x,t)\\
\psi_R(x,t)
\end{pmatrix}
\end{equation}
where $L, R$ denote the left and right part of the wavefunction respectively.

The walk may be initialized at the origin which happens to be a boundary condition for the system. 
In that case, we have $\psi_L(0,0)=a_0$ and $\psi_R(0,0)=b_0$; $a_0^2+b_0^2=1$. This ensures that $\psi_L(x \neq 0, 0)=\psi_R(x \neq 0, 0)=0$. As we have taken symmetric walk 
therefore, here, $a_0=\frac{1}{\sqrt{2}}, b_0=\frac{i}{\sqrt{2}}$.

The unitary operator we have chosen is Hadamard operator $H$ that acts on the chirality state. $H$ is given as
\begin{equation}
 H=\frac{1}{\sqrt{2}}\begin{pmatrix}
 1 & 1\\
 1 & -1
 \end{pmatrix}
\end{equation}

After the action of Hadamard coin the Translation operator $T$ acts in the following way:

\begin{equation}
\begin{aligned}
    T\ket{x,L} & = \ket{x-1,L}\\ 
    T\ket{x,R} & = \ket{x+1,R}
\end{aligned}
\label{eq4}
\end{equation}

QRW is usually studied in two specific ways: 
\begin{itemize}
 \item Without a detector : This allows free propagation of the walker on either side of origin. 
 This corresponds to an Infinite Walk (IW).
 \item With a detector : This is usually studied by placing a detector or absorbing boundary on 
 one side of the path of the walker. This restricts free propagation on one side and therefore
 leads to Semi Infinite Walk (SIW).
\end{itemize} 
Vis-a-vis this usual SIW there is another variation called the Quenched Quantum Walk (QQW), where the detector 
is initially placed at a site and then removed from the path of the walker after a certain time interval. 
In all the above cases with detector, we already know that 
presence of a detector at a certain site $x_D$ in the path of the walker affects the occupation probabilities 
of sites in the following way : If the walker reaches $x_D$ with probability $p_{x_D}$ and 
the detection probability of the detector is $p_D$, then the absorption probability would obviously be $p_{x_D}p_D$.

However, in our study, we have chosen that, whenever the walker reaches the detector, it is detected 
with probability unity, i.e., $p_D=1$. 
In addition to this, here the detector has a movement 
which we assume to be a hopping. The detector is initially placed at $x_D$. It then hops to a site $(x_D)_1=x_D+s$ after making $n$ number of detections. 
Therefore, if the first detection at $x_D$ is made at time $t$, 
the first detection at $(x_D)_1$ will be at time $t_1=t+(2n+s)$. When it completes $n$ detections at $(x_D)_1$ it moves 
further to $(x_D)_2=(x_D)_1+s$ and detects in the same manner. 

From this point of view, we can also say that there must be some 
kind of velocity associated with the detector. This can also be interpreted from a different point 
of view in case of experimental arrangements as discussed earlier. A detector at $x_D$ may lose its efficiency 
to detect after a finite number of detections. In 
that case, it may be replaced by a new one. In our case the new one is introduced at $(x_D)_k=x_D+ks$, where 
$k=1,2,3$ and so on for $t_k=t+k(2n+s)$. Note that replacing the old detector by a new one at the same place does not lead to any interesting behavior and therefore, we haven't shown that here. It is also worth mentioning that the backward movement of the walker, when $(x_D)_k=x_D-ks$ leads to probability distribution at sites very similar to SIW with a leftward shift and therefore it is not shown. 

The presence of the detector will modify the usual propagation of the walker and therefore occupation probabilities of sites
(probability distribution function) will be altered. It should be mentioned here that our usual notation for occupation probabilities  
$f(x,t)$ should be modified here. As $n$ and $s$ are the two important parameters 
responsible for the alteration of the occupation probabilities from now on, we will use the notation $f(x,t;n,s)$ or rather a shorter notation $f_{ns}$ to denote the probability distribution function. Also from now on, we will call our walk as Moving Detector Quantum Walk (MDQW). It is to be noted here that according to the measurement scheme proposed in \cite{goswami2}, here $\sum_x f(x,t;n,s)=1-p_a$ where $p_a$ is the
probability that it was absorbed earlier. Although we can choose the measurement scheme in two ways, one by normalizing the total probability over sites after each step of absorption/detection, and the other without normalizing, here we are choosing the second scheme. This has been done as we are interested in the actual enhancement of the probability at sites after changing the position of the detector, which does not happen due to the renormalization. In \cite{goswami} too, the scaling forms of several quantities have been studied without renomalization for the same reason.


\section{Results}
To compare our MDQW to the earlier versions of quantum walk, 
we first present a few snapshots of the occupation probabilities of sites for $t=1000$ for IW, SIW 
and MDQW. As we know from the definition of our walk, here we have two parameters, $n$ and $s$. In the subsequent part of this work, we will 
use a notation to identify the number of detection and amount of shift/hop as $nDsS$. For example, $2D6S$ 
represents that there will 
be $2$ detections at each position of a detector and amount of hop is $6$. If the detector is 
initially at $x_D$, the walker is not allowed to go beyond $x_D$. The shift/hop of the walker allows 
the walker to move beyond $x_D$, as then the detector has moved to $x_D+ks$ further on that side thereby allowing the 
walker to cross $x_D$. We expect three limiting cases here as Follows:
\begin{itemize}
    \item For $x_D >> 0$, the walker is not at all detected by the detector for long. This will lead to IW behavior.
    \item With $n$ large, it is obvious that the detector has to stay at a particular position for long and therefore it can be approximated to SIW.
    \item For $s \rightarrow \infty$, the situation is as if the detector will detect and then it will be removed from its position. This should resemble QQW.
\end{itemize}
In the following subsections we will try to see these limiting cases along with other important results.

\subsection{MDQW with variation of $n$}
\begin{figure}[h!] 
\begin{center}
\includegraphics[angle=-90, trim = 0 0 0 0, clip = true,width=0.75\linewidth]{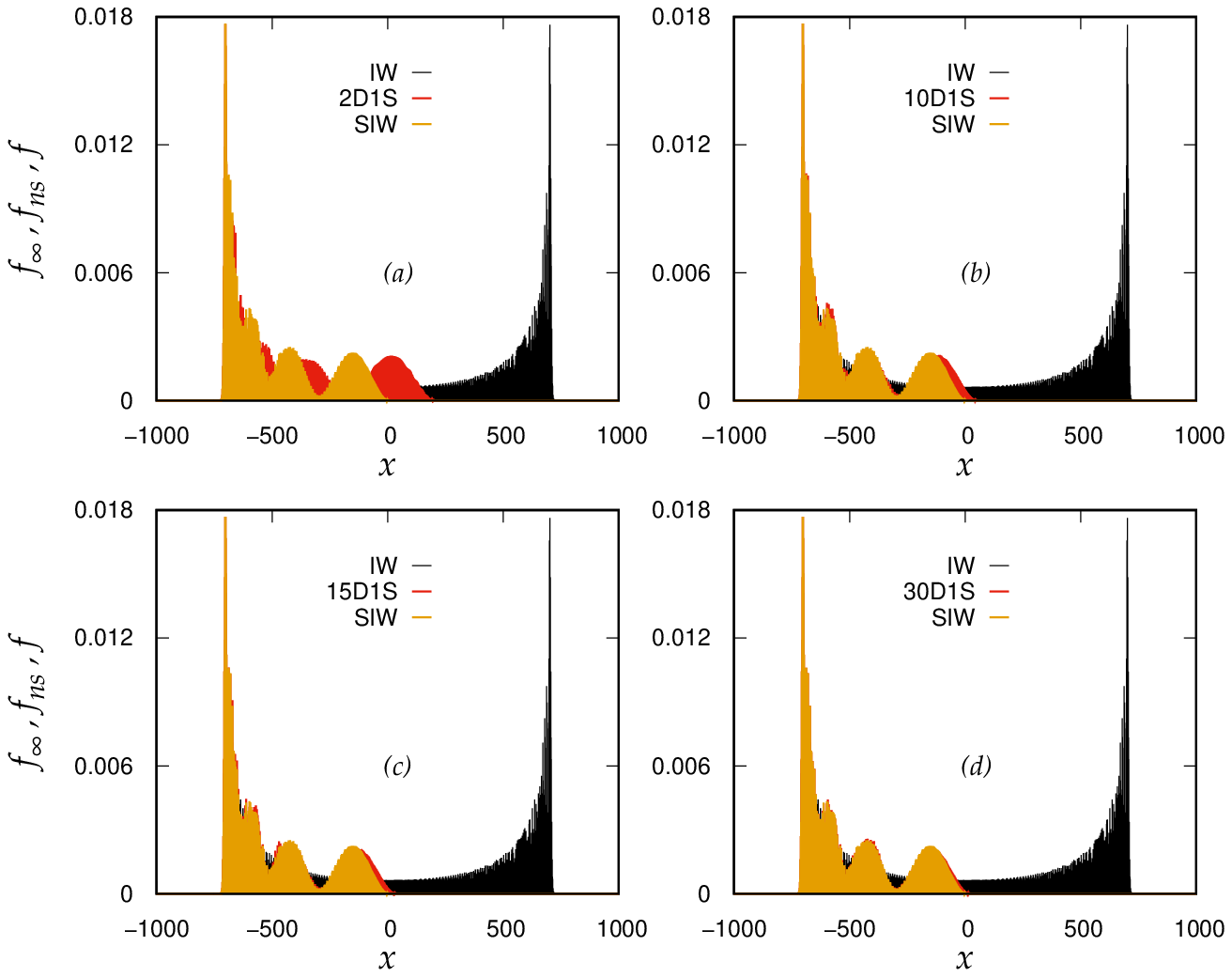}
\caption{Occupation Probability snapshots for a MDQW (symmetric) for $t=1000$ for (a) 2D1S, (b) 10D1S, (c) 15D1S and (d) 30D1S. In all the plots the SIW and IW cases are shown for comparison. Here $f_{\infty}, f_{ns}$ and $f$ are the occupation probabilities for IW, MDQW and SIW respectively. As $n$ increases, the approach of MDQW towards SIW behavior is clearly visible.}
\label{Multi}
\end{center}
\end{figure}

The occupation probabilities of sites for IW, MDQW and SIW, denoted as $f_{\infty}, f_{ns}$ and $f$ respectively are shown in form of snapshots in Fig. \ref{Multi}. In all the cases here, the amount 
of shift $s=1$ and the number of detections are $n=2, 10, 15, 30$ respectively with the 
detector initially placed at $x_D=10$. As here, the amount of shift is $1$, a small number of detection allows more spilling of the 
probability distribution function compared to a large number of detection.  
It is clear that as 
the number of detection $n$ increases, the occupation probability becomes more and more similar to the SIW occupation probability, as expected.

\begin{figure}[h!]
\begin{center}
\hspace*{-0.1in}
\includegraphics[angle=-90, trim = 0 0 0 0, clip = true,width=0.65\linewidth]{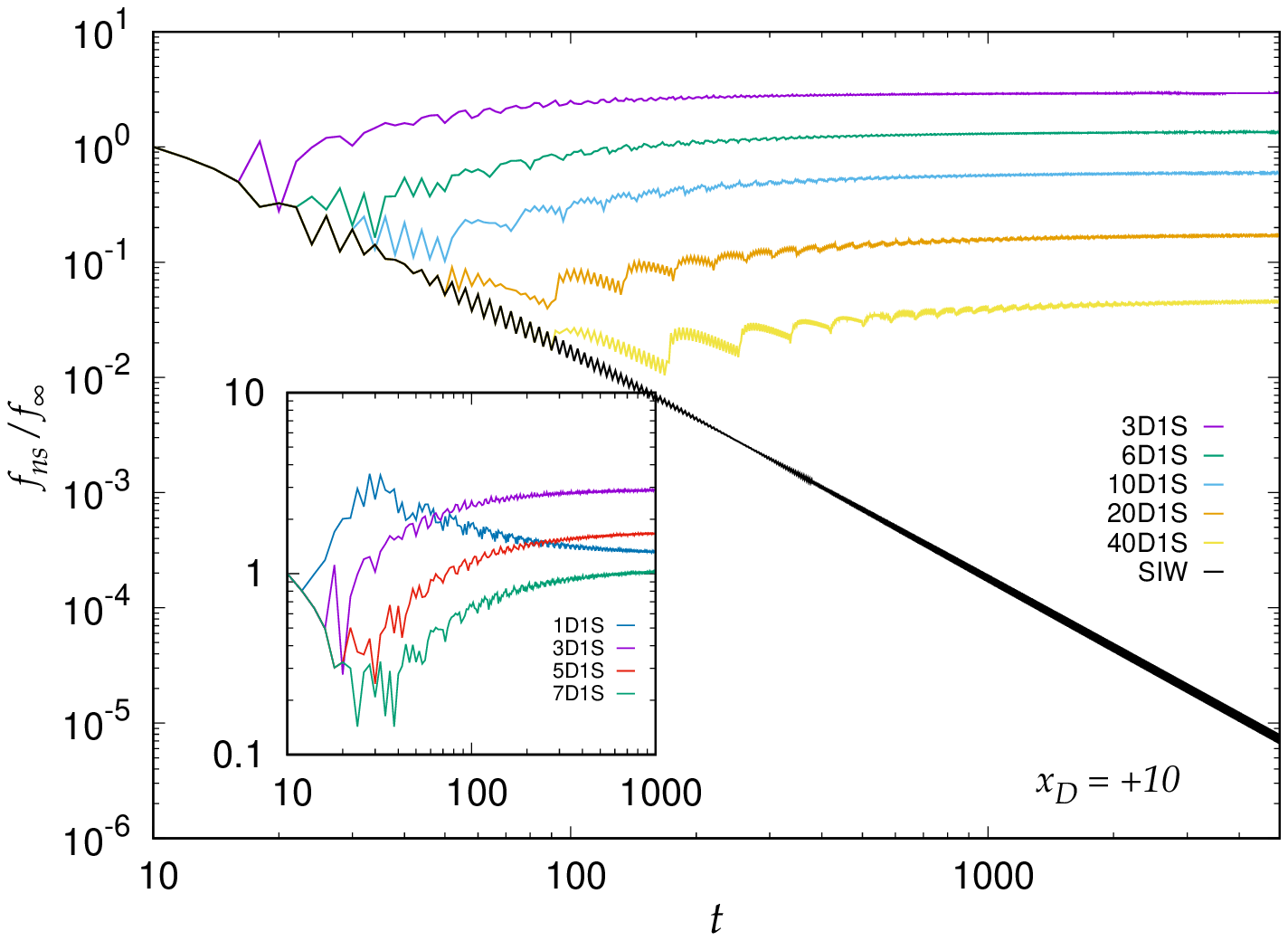}
\caption{Ratio of occupation probability of MDQW to that of IW, i.e., $\frac{f_{ns}}{f_{\infty}}$ as a function of $t$ is shown for $nD1S$ for $x_D=10$. For MDQW, for each $n$ the ratio approaches a saturation value. The same is shown for SIW for comparison. The small $n$ behavior of $\frac{f_{ns}}{f_{\infty}}$ is shown in the inset. For $1D1S$ the behaviour is different from other cases which is explained in the text.} 
\label{ff0_nD1S}
\end{center}
\end{figure}

\begin{figure}[h!]
\begin{center}
\hspace*{-0.1in}
\includegraphics[angle=-90, trim = 150 0 0 0, clip = true, width=0.85\linewidth]{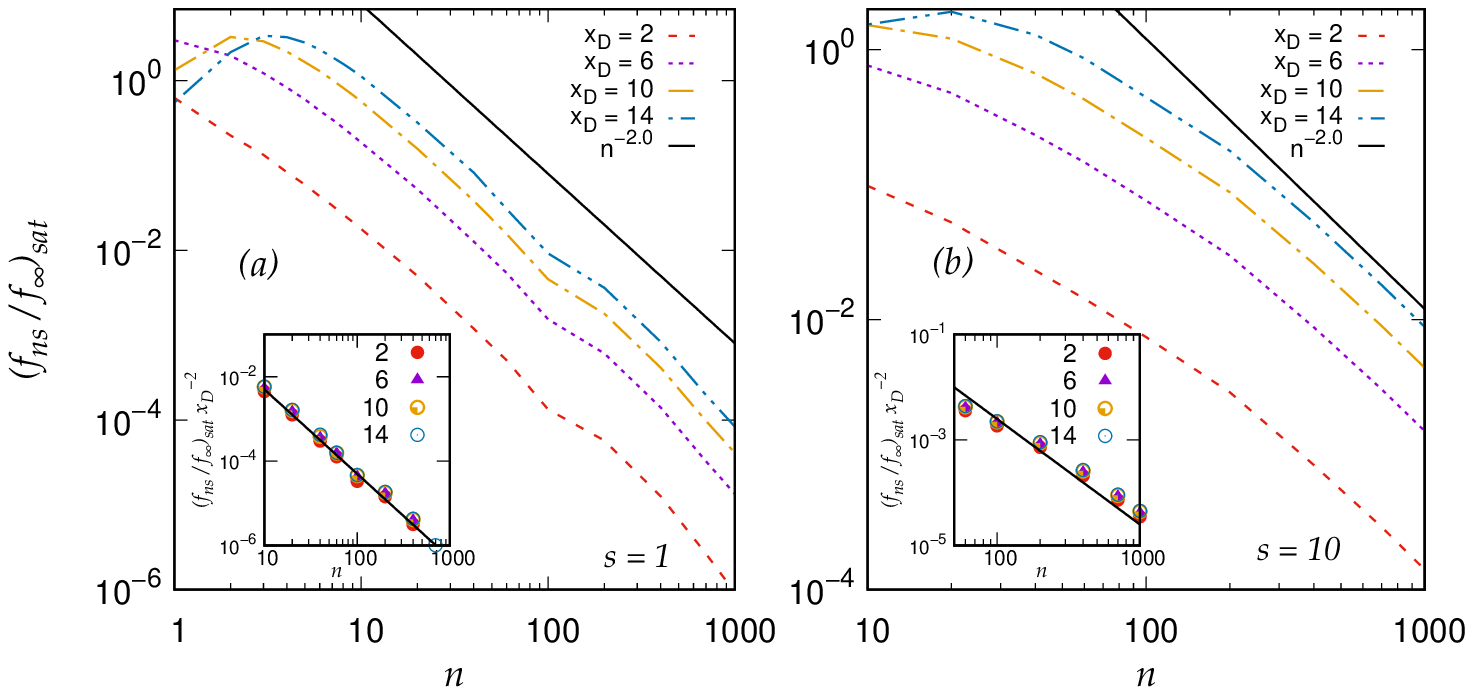}
\caption{$\left(\frac{f_{ns}}{f_{\infty}}\right)_{sat}$ against numbers of detection $n$ for $x_D=2, 6, 10, 14$ in (a) for $s=1$, (b) for $s=10$. A $n^{-2}$ behavior is observed for large $n$. In the respective insets, data collapses are shown (again for large $n$), the exponents being $2.04$ with an error bar of $0.005$.}
\label{ff0_sat_vs_n}
\end{center}
\end{figure}
We now wish to see this in more detail. For this purpose, the ratio of occupation probability of a site for a MDQW, i.e. $f_{ns}$ (a shorter notation for $f(x,t;n,s)$)
and that of an IW, i.e., $f_{\infty}$ is shown against $t$ for $nD1S$ where the initial position of the detector $x_D=10$ in Fig. \ref{ff0_nD1S}. In the same Fig. the corresponding ratio is shown for a SIW where the 
detector is not replaced or removed. 
As the initial position of the detector is $x_D=10$, it is obvious that up to $t=x_D=10$ (here), 
the presence of the detector cannot affect the walker. Beyond $x_D$ the occupation probabilities of 
sites for the walker starts getting affected by the detector and therefore the ratio 
starts decreasing below $1$. Here the number of detection $n$ plays a very important role. It is to 
be mentioned here once again that $n$ is actually related to the efficiency of the detector. Here $n$ has shown to vary from $1$ to $40$. It is clear that as long as the detector 
is placed at initial position $x_D$, the plot is same as SIW. If we look closely, the pattern is same as SIW upto 
$t=2n+x_D$. After this time the detector will hop to $(x_D)_1$. Thus, when the walker reaches $x_D$ next, there is no detection, as the detector is already moved to a new position. 
The ratio for site $10$ will increase therefore. As the walker 
reaches the detector again the tendency to decrease starts. The hopping of the detector leads to a kind of periodic 
behavior as seen from the plot. However, after a sufficiently long time the ratio of $f_{ns}/f_{\infty}$ at site $10$ 
saturates. In the inset of Fig. \ref{ff0_nD1S}, the same $f_{ns}/f_{\infty}$ behavior is shown when numbers of detection 
are small. For $1D1S$, after the walker gets detected by the detector once, the detector hops to the new position. Therefore, for $x_D=10$, after $t=12$, the ratio starts to increase. This increase is much more prominent compared to other cases as here the detector has detected only once. After a certain time it decreases because the effect of the detector is never very small as after each detection it shifts only by one step. The non-monotonic behaviour for $1D1S$ is actually the signature of competing behavior of effect of $n$ and $s$. 
It is clear from all the plots that the ratio ultimately approaches a saturation value $\left(\frac{f_{ns}}{f_{\infty}}\right)_{sat}$. For low $n$ 
values at a large $t$ having a very high $\left(\frac{f_{ns}}{f_{\infty}}\right)_{sat}$ much greater than $1$ suggests that the walker will try to 
compensate for the sites not reached.  
In Fig. \ref{ff0_sat_vs_n} (a) and (b) $\left(\frac{f_{ns}}{f_{\infty}}\right)_{sat}$ is shown as a function of number of detection $n$ when 
the detector is initially placed at $x_D$ for $s=1, 10$ respectively. The plots show a behavior $n^{-\alpha}$ for large $n$ for all the cases. 
\begin{figure}[h!]
\begin{center}
\includegraphics[angle=-90, trim = 0 0 0 0, clip = true,width=0.65\linewidth]{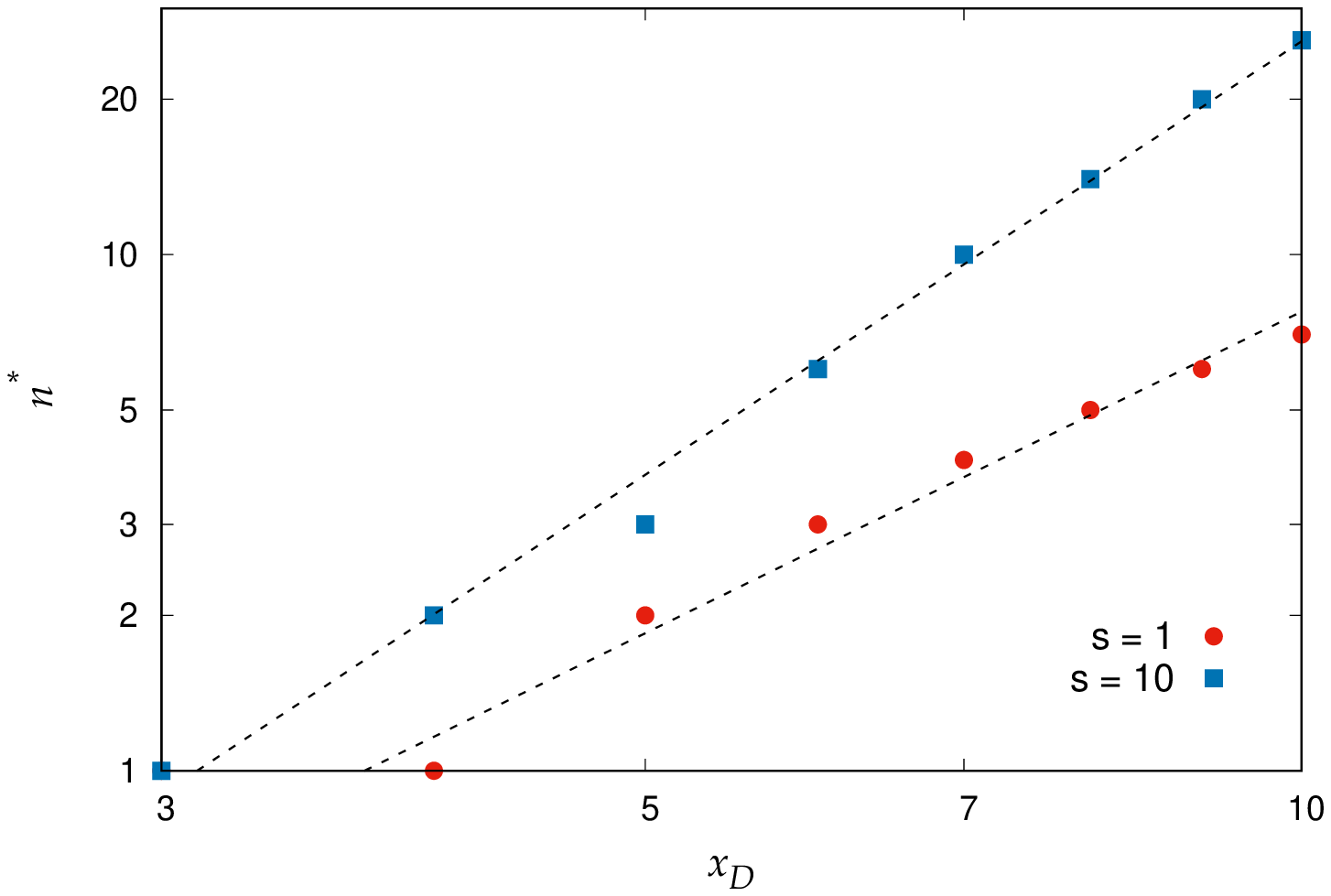}
\caption{The variation of $n^*$ against $x_D$ is shown for $s=1, 10$. Power law behaviour is observed; $n^* \sim x_D^\xi$. For $s=1$ and $10$, $\xi=2.0$ and $2.8$ respectively.}
\label{nstar_n}
\end{center}
\end{figure}
The exponent $\alpha$ here has been 
observed to be nearly equal to $2$ for any fixed $s$. However, for large hopping $s$, small $n$ affects the system 
to a lower degree and therefore the ratio stays close to $1$ even for large $n$. This effect is more prominent for higher $s$ but we have not shown it here. The collapsed data are shown in the insets of Fig. \ref{ff0_sat_vs_n}. Therefore, it may be written that for large $n$
\begin{equation}
    \left(\frac{f_{ns}}{f_{\infty}}\right)_{sat} \sim \frac{x_D^2}{n^2}
    \label{eq}
\end{equation}
As the initial position of the detector $x_D$ is more towards right (means the value of $x_D$ is increased), the walk remains unaffected for longer times. Therefore, $\left(\frac{f_{ns}}{f_{\infty}}\right)_{sat}$ should increase with $x_D$. It is also clear that with increase in number of detection $n$, the walk gets affected increasingly and $\left(\frac{f_{ns}}{f_{\infty}}\right)_{sat}$ should decrease with $n$. 
However, the exact scaling behaviour of Eq. \ref{eq} is not obvious.
It should be mentioned that there is a certain 
number of detections $n^*$ for each fixed $x_D$ and $s$ beyond which $\left(\frac{f_{ns}}{f_{\infty}}\right)_{sat}$ can never attend $1$. 
The plot of $n^{*}$ against $x_D$ shows power law fit as observed in Fig. \ref{nstar_n}. This variation pf $n^*$ supports the nature observed in variation of $\frac{f_{ns}}{f_{\infty}}$ against $t$. This is already indicated in Fig. \ref{ff0_nD1S} as for small $n$ if we can wait for sufficiently long time, the system can qualitatively approach IW but not in case of large $n$.

\subsection{MDQW with variation of $s$}
Let us now move on to the study of MDQW from the aspect of another parameter which we can control, i.e., the amount of shift/hop $s$. We now fix the 
number of detection $n$ and vary $s$. 
\begin{figure}[h!]
\begin{center}
\includegraphics[angle=-90, trim = 0 0 0 0, clip = true,width=0.75\linewidth]{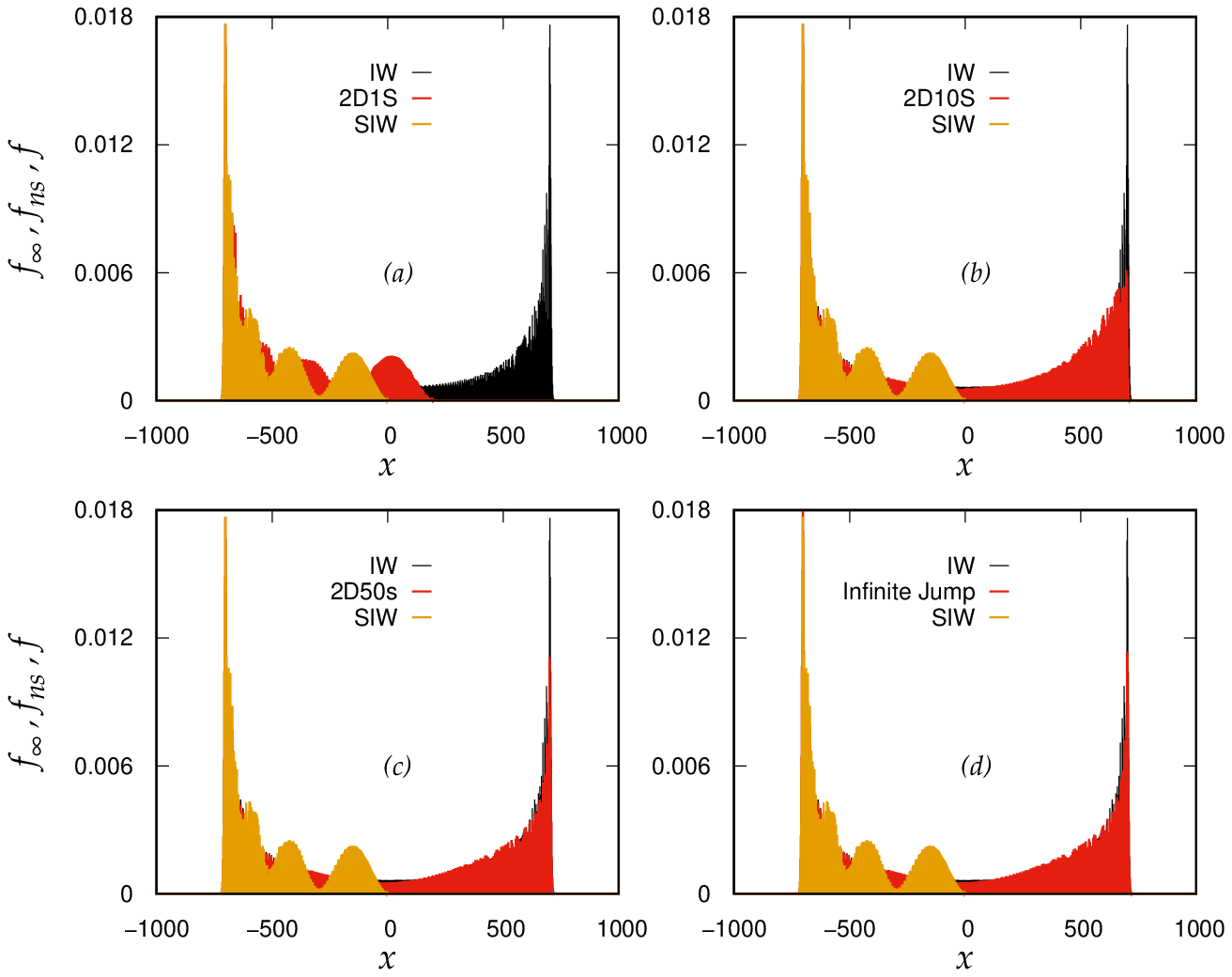}
\caption{Occupation Probability snapshots for a MDQW (symmetric) for $t=1000$ for (a) 2D1S, (b) 2D10S, (c) 2D50S and (d) 2DIJ (IJ denotes Infinite Jump). In all the plots the SIW and IW cases are shown for comparison. As $s$ increases, the qualitative approach towards IW behavior is visible. However, this is small $n$ behavior as discussed in the literature. Note that $x_D=10$ here.}
\label{2DsS_Multi}
\end{center}
\end{figure}
\begin{figure}[h!]

\begin{center}
\includegraphics[angle=-90, trim = 0 0 0 0, clip = true,width=0.65\linewidth]{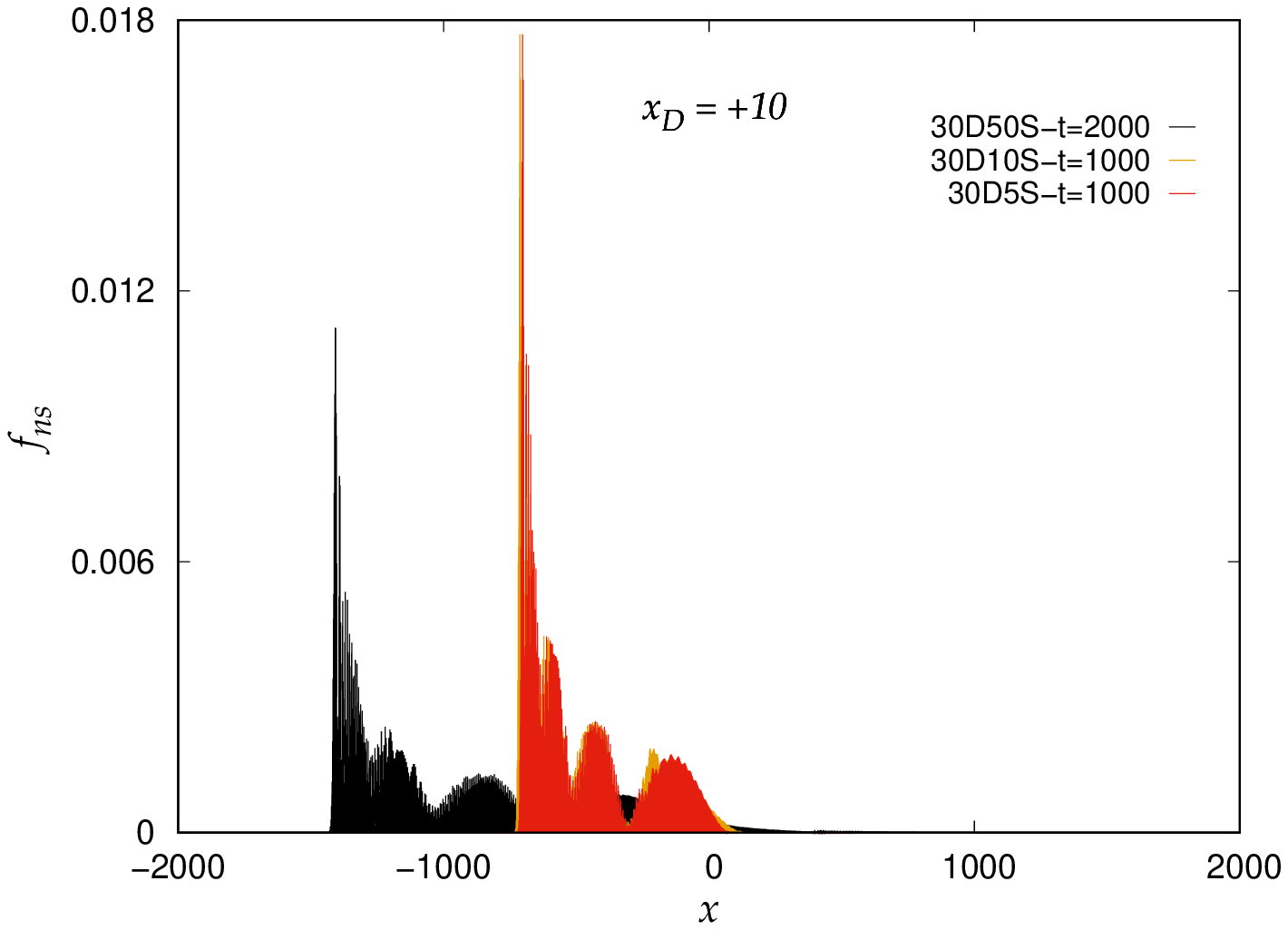}
\caption{Occupation Probability snapshots for a MDQW (symmetric) for 30DsS for $t=1000$ and $t=2000$ for $x_D=10$. For $n$ as large as $30$, the spilling is seen to be considerably low. Even for a longer time snapshot, there is no significant spilling towards the right side where initially the detector was placed.}
\label{30DsS_single}
\end{center}
\end{figure}
In Fig. \ref{2DsS_Multi}, we have shown the snapshots for $2DsS$. It is seen that as we increase the amount of shift $s$, the tendency of spilling of the probability distribution beyond the initial position of the detector increases as the detector has already moved then. It is to be noted that here the longest shift/hop is taken to be an Infinite Jump, denoted as IJ. This naively suggests that there is a possibility that we may achieve the probability distribution of IW, at least qualitatively, in this case if we can make the amount of shift very high. 
However, from Fig. \ref{30DsS_single}, it is clear that if the number of detection is more, the spilling is considerably low and even for a longer time snapshot, it is hard to see any significant spilling over the right side where initially the detector was placed. 
\begin{figure}[h!]
\begin{center}
\includegraphics[angle=-90, trim = 100 0 0 0, clip = true,width=0.85\linewidth]{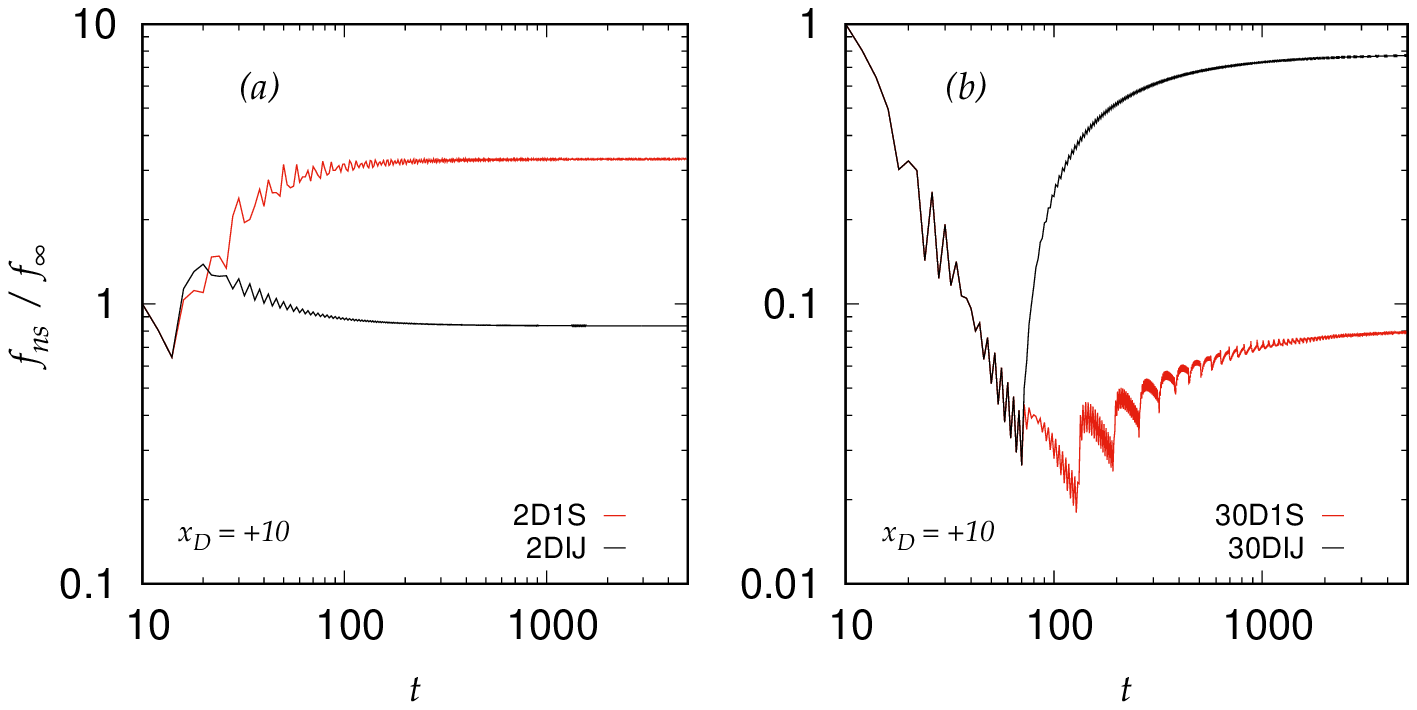}
\caption{Variation of $\frac{f_{ns}}{f_{\infty}}$ against $t$ for (a) 2D1S, 2DIJ and (b) 30D1S, 30DIJ (IJ denotes Infinite Jump). The individual curves approach different saturation values denoted as $\left(\frac{f_{ns}}{f_{\infty}}\right)_{sat}$.}
\label{ff0_t_varyshift}
\end{center}
\end{figure}
This means the more number of detection affects the nature of the walk significantly.\\ 

The ratio $\frac{f_{ns}}{f_{\infty}}$ for these sets show some interesting facts. When $n$ is small, for small $s$, the ratio saturates to a sufficiently high value above $1$. The saturation value is realized as $\left(\frac{f_{ns}}{f_{\infty}}\right)_{sat}$. For small $n$, high value of $\left(\frac{f_{ns}}{f_{\infty}}\right)_{sat}$ indicates that the walker tries to move more towards the side where it was unable to  reach before. 
\begin{figure}[h!]
\begin{center}
\includegraphics[angle=-90, trim = 0 0 0 0, clip = true,width=0.65\linewidth]{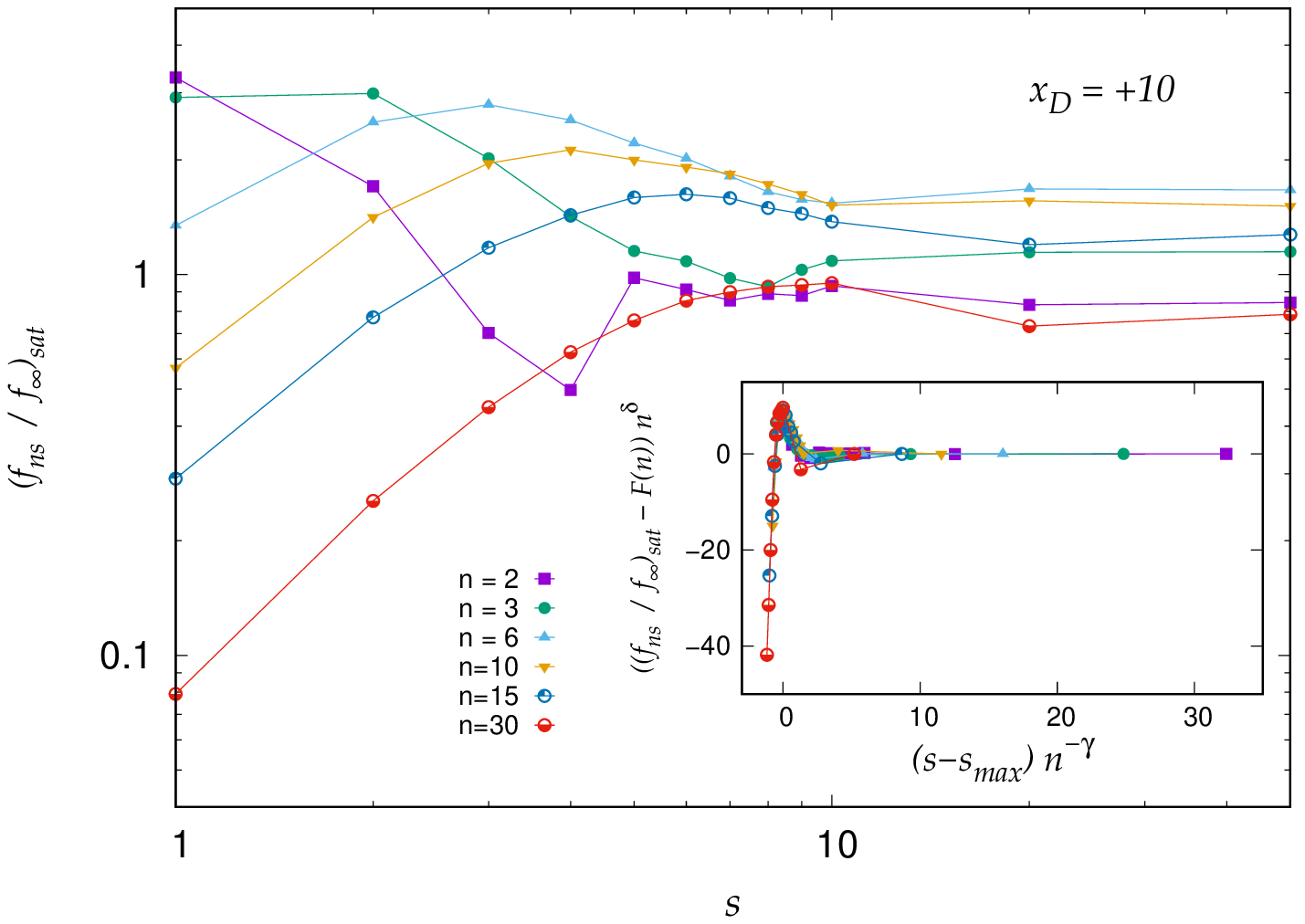}
\caption{$\left(\frac{f_{ns}}{f_{\infty}}\right)_{sat}$ against amount of shift $s$ for $n=2,3,6,10,15,30$. For small $n$, $\left(\frac{f_{ns}}{f_{\infty}}\right)_{sat}$ first decreases, then increases and finally approaches $F(n)$. For high $n$, it increases and then decreases slightly to approach $F(n)$. The maximum value of the individual curves for different $n$ occur at different $s_{max}$. It is understood that the curves approach $F(n)$ for $s>>s_{max}$. The data here is shown for $x_D=10$. Data collapse for the same is shown in the inset with $\gamma=0.6$ and $\delta=1.2$.}
\label{ff0sat_s}
\end{center}
\end{figure}
As we increase $s$, the saturation first decreases below $1$ and then again starts increasing and finally for high $s$
approaches a value below $1$. This is shown in Fig. \ref{ff0_t_varyshift} (a). For a very large number of detection $n$, as can be seen from Fig. \ref{ff0_t_varyshift} (b) for $n=30$, there is more and more  absorption and the ratio 
saturates to a value less than $1$. With increase in amount of shifts $s$, there is increase in $\left(\frac{f_{ns}}{f_{\infty}}\right)_{sat}$. However, it is checked that
beyond $n=21$ whatever be the value of $s$, the ratio can never attain $1$.
The saturation value of $\frac{f_{ns}}{f_{\infty}}$, i.e., $\left(\frac{f_{ns}}{f_{\infty}}\right)_{sat}$ is shown as a function of amount of shifts $s$ for $x_D=10$ in Fig. \ref{ff0sat_s} for different $n$.  For high $n$, the value of $\left(\frac{f_{ns}}{f_{\infty}}\right)_{sat}$ first increases, reaches a maximum at $s_{max}$ and then decreases slightly to approach a value $F(n)$. However for smaller $n$ values $\left(\frac{f_{ns}}{f_{\infty}}\right)_{sat}$ decreases from it maximum value at $s=s_{max}$, then increases to reach a saturation $F(n)$. Therefore, $F(n)$ can be understood as the value of $\left(\frac{f_{ns}}{f_{\infty}}\right)_{sat}$ when $s>>s_{max}$. A scaling behavior for $s_{max}$ is observed where $s_{max} \sim n^{0.77}$. Using this $s_{max}$ and $F(n)$, we found data collapse for $\gamma=0.6$ and $\delta=1.2$. The data collapse is shown in the inset of Fig. \ref{ff0sat_s}. The approximate behavior of $\left(\frac{f_{ns}}{f_{\infty}}\right)_{sat}$ as obtained from the collapsed data is 
$F(n)+n^{-\delta}\mathcal{F}\left((s-s_{max})n^{-\gamma}\right)$. In our study the behavior of MDQW for $s >> s_{max}$ is closely related to the removal of detector for QQW as in \cite{goswami}. We have checked that the removal time of the detector $t_R$ as in \cite{goswami} and our parameters $n$ and $s$ are related as $t_R = x_D + 2(n-1)$. The corresponding saturation values are in good agreement with each other. 



\subsection{Probability distribution ratio for $x \neq x_D$}
\begin{figure}[h!]
\begin{center}
\includegraphics[angle=-90, trim = 0 0 0 0, clip = true,width=0.75\linewidth]{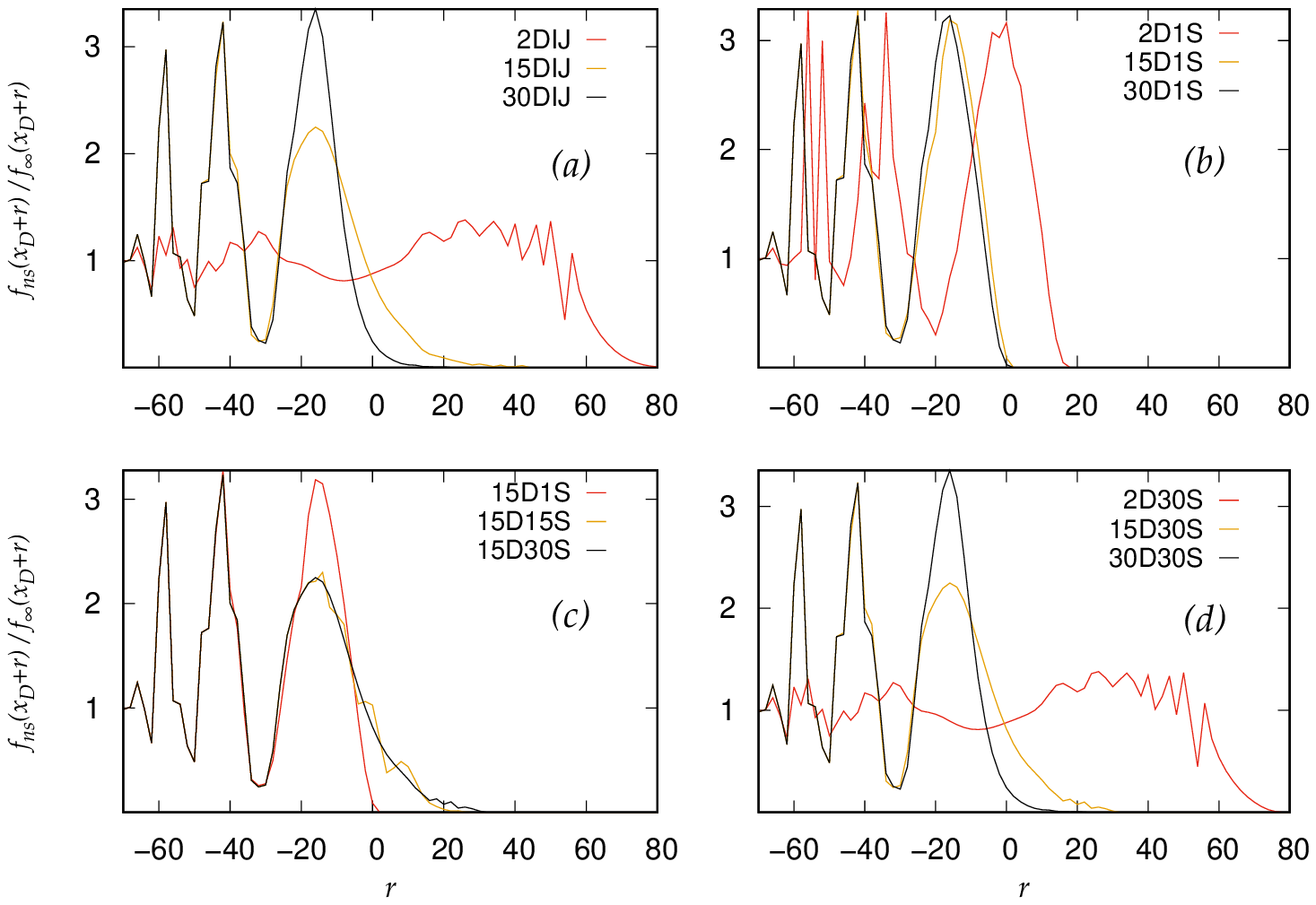}
\caption{Variation of $\frac{f_{ns}(x_D+r)}{f_{\infty}(x_D+r)}$ against $r$ for MDQW for (a) nDIJ, (b) nD1S, (c) 15DsS and (d) nD30S. The plots corresponding to (a) and (d) are almost similar.}
\label{r_variation}
\end{center}
\end{figure}

Now let us try to see what happens for other sites $x \neq x_D$. It is clear that when we consider infinite jump 
the behavior of the walk will be similar to QQW. In Fig. \ref{r_variation} (a) it is shown. 
The ratio of the occupation probabilities $\frac{f_{ns}(x_D+r)}{f_{\infty}(x_D+r)}$ show some interesting behavior as follows:
\begin{itemize}
   \item For $r<0$, several peaks are observed with the peak values much greater than $1$. For large negative $r$, the ratio is $1$ implies that the MDQW and IW behave there in the same way. 
    \item for $r>0$, 
    $\frac{f_{ns}(x_D+r)}{f_{\infty}(x_D+r)}$ 
    goes to zero at a finite value of $r$. The decay is 
smooth for large values of $n$ whatever be the value of $s$. For large $n$ and small $s$ the ratio drops off sharply close to $r=0$. For small $n$, however, the decay is dependent on $s$.  When $s$ is small, its behavior resembles any high $n$ behavior as shown in Fig. \ref{r_variation} (b), but for high $s$, the decay is accompanied by small oscillations. Even when 
the hop is not infinite but as large as $30S$, the behavior is the same as IJ as is shown in the \ref{r_variation} (d) plot. 
\end{itemize}

As is evident from Fig. \ref{r_variation} (a), (b), (d) that fixing $s$ and changing $n$ would lead to a shift in the position of the peaks and would also lead to a change in their heights. For small $s$, the effect over the height is not prominent. As we start increasing $n$, at the beginning the effect of shift is strong but when $n$ is large enough the position of peaks becomes somewhat fixed. Keeping $n$ fixed and changing $s$ affect mostly the peak heights of the first peaks although their positions do not change, except from some broadening as in seen in Fig. \ref{r_variation} (c). The approximate form of the function can be written as:
\begin{equation}
    \frac{f_{ns}(x_D+r)}{f_{\infty}(x_D+r)} \sim A+Br^{\nu}\sin(r^\beta)\exp({-r^2/D})
    \label{eqn5}
\end{equation}
The corresponding parameters are shown in Table \ref{table1}.

\begin{table}[h!]
\begin{center}
\begin{tabular}{|c|c|c|c|c|c|c|}
\hline
      n  &  s  & A & B & $\nu$ & $\beta$ & D \\
     \hline
     2 & 1 & 1.05 & 0.5 & 0.2 & 0.456 & 3 $\times 10^8$ \\
     & 15 & 1.05 & 0.06 & 0.2 & 0.4 & 3 $\times 10^7$ \\
     & 30 & 1.05 & 0.06 & 0.2 & 0.4 & 3 $\times 10^8$ \\
     \hline
     15 & 1 & 1.7 & 0.6 & 0.19 & 0.54 & 5 $\times 10^5$ \\ 
     & 15 & 1.7 & 0.7 & 0.19 & 0.54 & 7 $\times 10^4$ \\
     & 30 & 1.7 & 0.7 & 0.19 & 0.54 & 6 $\times 10^4$ \\
     \hline
     30 & 1 & 1.7 & 0.75 & 0.15 & 0.6 & 3 $\times 10^5$ \\
     & 15 & 1.7 & 0.86 & 0.15 & 0.598 & 2 $\times 10^5$ \\
     & 30 & 1.7 & 0.86 & 0.15 & 0.598 & 2 $\times 10^5$ \\
     \hline
\end{tabular}
\end{center}
\caption{Approximate values of the parameters used in Eq. \ref{eqn5} }
\label{table1}
\end{table}

As there is variation in $r$, it is important to study the correlation of occupation probabilities in $r$. The correlations for MDQW and IW are defined as:
\begin{equation}\label{eq6}
    \begin{aligned}
        g_{ns}(x_D+r) &= f_{ns}(x_D+r)f_{ns}(x_D) \\
        g_{\infty}((x_D+r) &= f_{\infty}(x_D+r)f_{\infty}(x_D)
    \end{aligned}
\end{equation}
respectively.
\begin{figure}[h!]
\begin{center}
\includegraphics[angle=-90, trim = 100 0 0 0, clip = true,width=0.85\linewidth]{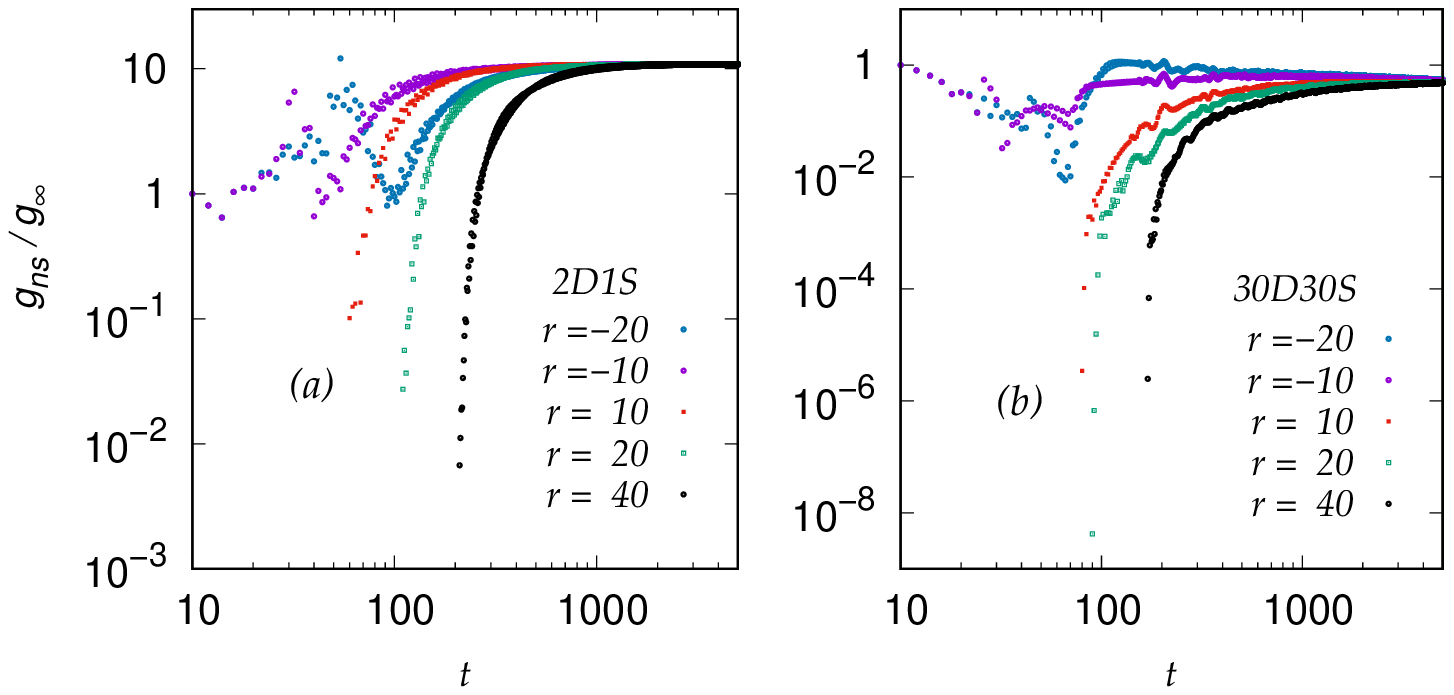}
\caption{Correlation ratio $\frac{g_{ns}}{g_{\infty}}$ as a function of $t$ for $r = -20, -10, 10, 20 $ and $40$ for (a) 2D1S and (b) 30D30S.}
\label{gg0_t_varyshift}
\end{center}
\end{figure}
The ratio of these two $r$ correlations as in Eq. \ref{eq6}, i.e., $\left(\frac{g_{ns}}{g_{\infty}}\right)$ is an important quantity. For different combinations of $n$ and $s$, the behavior of this correlation ratio for different $r$ is studied as a function of $t$ in Fig. \ref{gg0_t_varyshift}. The observed specific behavior are as follows: 
\begin{itemize}
    \item For small $n$: 
    \begin{itemize}
        \item In this case if $s$ is small, the ratio saturates above unity. For sufficiently negative $r$, it approaches the saturation from well below that value. As $r$ is increased, the approach towards saturation starts from a value closer to the saturation. However, as $r>0$ again the ratio approaches saturation from well below of that value. The reason that saturation is much higher than $1$ is connected to the fact that after the detection the system has a high tendency to go towards the side where the walker was initially placed. As $s$ is low the system do not find time to relax and therefore the tendency persists. This is shown in Fig. \ref{gg0_t_varyshift} (a). 
        \item For large $s$, the correlation approaches the saturation below $1$ for any $r$ from above. This is because the system now gets enough time to relax and it tries to approach IW (Plot not shown).
    \end{itemize}
    \item For large $n$: 
    \begin{itemize}
    \item If $s$ is small the correlation ratio saturates well below $1$. For $r < 0$ the approach towards saturation is from above and for $r > 0$ from below. The saturation decreases with $n$, as larger $n$ means greater absorption (Plot not shown). 
    \item For larger $s$ the correlation ratio saturates close to unity as in that case the system tries to compensate for the occupation probabilities for IW. This is shown in Fig. \ref{gg0_t_varyshift} (b). 
    \end{itemize}
\end{itemize}
\section{Summary}
In this work we have studied the detailed effect of a detector in a quantum
system from the aspect of Quantum Random Walk. As there are enormous number of experiments on QRW going on in recent years, the role and limitations of detectors is becoming a very important aspect as indicated in several studies. Like QQW, here also it is evident that the occupation probability 
of sites may be enhanced compared to IW by removing the detector and placing it at a different position. This is a purely quantum mechanical effect. However, this is not true always. If the efficiency of the detector, placed for example at a site towards right, is high, which means it can act longer, then the occupation probability cannot approach the IW picture on the right. It also depends on the initial position of the detector. For a particular initial condition the walker cannot go beyond $x_D$ up to a time $t_D=x_D+2n$. After that as the detector has hopped to another position, for $t>t_D$, the walker has increased freedom to move beyond $x_D$ but is again restricted by the new position $(x_D)_1$. In this way the walker gets its freedom towards right in steps. From Fig. \ref{Multi} and Fig. \ref{2DsS_Multi}, it is obvious that most of the contributions to
$x_D$ and beyond come from the density of walkers close
to it. This is because the occurrence probabilities far away from $x_D$ are
not much affected by the removal of the detector. At
larger times after the replacement of the detector, the occupation probability distribution approaches the IW picture. This effect is prominent for small $n$ and very large $s$. First the local hill like structures closest to $x_D$ smooths out and subsequently further parts are affected. However, for large $n$ and small $s$, the MDQW approaches SIW. Another important limiting case is when we have infinite jump for the detector, i.e., $s \rightarrow \infty$ which give results similar to QQW. It is also checked that as $s>>s_{max}$, the $F(n)$ values obtained as in Fig. \ref{ff0sat_s} are in good agreement to the results obtained in \cite{goswami}.

The scaling behavior of $\left(\frac{f_{ns}}{f_{\infty}}\right)_{sat}$ is an important quantity which is found to scale as $\frac{x_D^2}{n^2}$ here. The $n^{-2}$ behavior is robust against variation of $x_D$ and $s$, although for large $s$ the behavior starts from larger $n$. For a fixed $x_D$ and $s$ there is a certain number
of detection $n^*$ beyond which the ratio $\left(\frac{f_{ns}}{f_{\infty}}\right)_{sat}$ goes below $1$ and cannot increase thereafter. The $n^*$ versus $x_D$ curve shows power law behavior ($n^* \sim x_D^\xi$; $\xi$ being $0.77$ here). Another important result is the scaling of $\left(\frac{f_{ns}}{f_{\infty}}\right)_{sat}$ when plotted as a function of $s$ with $n$ as a parameter. The behavior can be approximated as 
$F(n)+n^{-\delta}\mathcal{F}\left((s-s_{max})n^{-\gamma}\right)$ with $\gamma=0.6$ and $\delta=1.2$ as is shown in Fig. \ref{ff0sat_s}. 

The behavior of occupation probability ratio $\frac{f_{ns}}{f_{\infty}}$ at sites different from the initial position of the detector, i.e., at $x=x_D+r$ ($r$ may be positive or negative as already discussed in section III C) shows some definite behavior as is shown in Fig. \ref{r_variation}. This behavior can be approximated as 
$A+Br^{\nu}\sin(r^\beta)\exp({-r^2/D})$ and the variation of the parameters with $n$ and $s$ are studied. The corresponding correlation ratios are also shown to have certain interesting behavior.   

Our work involves a detector with $p_D=1$ and its replacement is realized in form of a movement as discussed. This work can be extended by studying the response of the system when $p_D$ is some definite function of time and thereby comparing it to the experimental data available, if any.  

\begin{center}
    \textbf{Acknowledgement}
\end{center}

AM acknowledges financial support from CSIR (Grant no. 08/0463(12870)/2021-EMR-I). AM and SG acknowledges the computational facility of Vidyasagar College.

\end{document}